\documentclass{./style/llncs}
\usepackage{hyperref}
\usepackage{tikz}
\usepackage{graphicx}
\usepackage{cite}
\usepackage[capitalize]{cleveref}
\begin{document}
\mainmatter              
\title{AutoML Segmentation for 3D Medical Image Data: Contribution to the MSD Challenge 2018}
\titlerunning{AutoML Segmentation for 3D Medical Image Data}  
%
\author{Oliver Rippel \and Leon Weninger \and Dorit Merhof}
\authorrunning{Oliver Rippel et al.} 
%
\tocauthor{Oliver Rippel, Leon Weninger, Dorit Merhof}
\institute{Institute of Imaging \& Computer Vision \\ RWTH Aachen University \\ Aachen, Germany\\
\email{rippel@lfb.rwth-aachen.de}}

\maketitle              

\begin{abstract}
Fueled by recent advances in machine learning, there has been tremendous progress in the field of semantic segmentation for the medical image computing community.
However, developed algorithms are often optimized and validated by hand based on one task only.
In combination with small datasets, interpreting the generalizability of the results is often difficult.
The Medical Segmentation Decathlon challenge addresses this problem, and aims to facilitate development of generalizable 3D semantic segmentation algorithms that require no manual parametrization.
Such an algorithm was developed and is presented in this paper. It consists of a 3D convolutional neural network with encoder-decoder architecture employing residual-connections, skip-connections and multi-level generation of predictions.
It works on anisotropic voxel-geometries and has anisotropic depth, i.e., the number of downsampling steps is a task-specific parameter.
These depths are automatically inferred for each task prior to training.
By combining this flexible architecture with on-the-fly data augmentation and little-to-no pre-- or postprocessing, promising results could be achieved.

The code developed for this challenge will be available online after the final deadline at: \url{https://github.com/ORippler/MSD_2018}
\end{abstract}
\section{Introduction}
Segmentation is one of the fundamental tasks of medical image computing \cite{Ayache2016}.
Here, recent advances in machine learning have yielded steadily-increasing performances of developed algorithms \cite{Shen2017}.
However, critical evaluation of further algorithmic advances in the field is limited by two facts \cite{Maier-Hein*2018}:

\begin{enumerate}
  \item Algorithms are often manually adapted and fine-tuned to the task at hand. Multi-task evaluation is seldomly performed.
  \item By changing the design paradigm to a data-driven one, algorithmic performance heavily depends on size and quality of the available datasets.
\end{enumerate}

Together, these two facts make it difficult to infer insights about the generalizability of newly proposed algorithms and to identify key advances.

The Medical Segmentation Decathlon (MSD) is a challenge addressing the above limitations by providing ten high-quality datasets to serve as the basis of multi-task algorithm design and evaluation \cite{Cardoso2018}.
Based on the given data, a multi-task algorithm has to be designed that can perform well across all individual tasks.
Manual adpatation of parameters to a given task is prohibited, i.e., every task-specific parameter of the designed algorithm has to be discerned automatically from the dataset by means such as cross-validation.

Our approach is based on a convolutional neural network (CNN) with an encoder-decoder architecture.
The details of the CNN, its automatic task-specific parametrization as well as training/inference procedures and pre- and postprocessing steps shall be given in the following.

\section{Methods}

\subsection{Pre-processing}
For pre-processing, task-specific resampling to the median voxel-spacing of the training dataset was performed.
This was based on the assumption that resampling to median voxel-spacing should maintain all structures/textures required to solve the given task.
Furthermore, median shapes were determined over all volumes for each task to derive model depths as well training patch-size.

\subsection{Model architecture}
For the model architecture, a 3D convolutional encoder-decoder architecture was developed based on the 3D U-Net \cite{Cicek2016}.
Apart from the skip-connections inherent to the U-Net, the model also employs residual-connections as well as a multi-level generation of segmentation predictions, which have both been shown to be beneficial for semantic segmentation of biomedical images \cite{Milletari2016, Dou2017, Chen2017}.
Similar to \cite{Isensee2018}, instance normalization is applied between convolutional and activation layers, as it has been hypothesized to be more stable to stochasticity introduced by small batch-sizes \cite{Isensee2018, Ulyanov2016}. 
The number of basis features is being set to 6 for all tasks.

The model has anisotropic depth, meaning that the number of downsampling steps performed in each dimension $d_i$ is given by the median task shapes in the following manner:

\begin{equation}
  d_i = min(\frac{shape_i}{2^{3}}, 4)
\end{equation}

The depth is limited to 4 for each dimension in order to facilitate training on memory-constrained GPUs.
This yields an anisotropic receptive field, with the largest possible being $[157,157,157]$ voxels at the lowest level of the encoder.
This is realized by setting strides as well as kernel size to 1 for every downsampling step that would reduce the number of voxels in the feature volume for any dimension below 8.
For the decoder, the upsampling kernels are chosen accordingly such as that spatial correspondence is maintained.
Furthermore, $max(depths) - 1 $ segmentation predictions are generated, ensuring that the gradient is propagated into deeper levels of the model.
By flexible determination of depths and receptive field sizes, the model should be well-suited to both isotropic as well as anisotropic voxel geometries.


\subsection{Model training}
For model training, the Adam optimizer \cite{Kingma2014} was chosen with an initial learning rate of 0.0005.
Learning rate was decayed by a factor of 0.984 every 100 steps, and total number of iterations were 30k for each task.

Due to GPU memory constraints, patch-wise training was performed.
Batch-size was 4, and depending on the task, multi-GPU training was utilized.
Patch-size was chosen to be $min([2^{3 + depth}],[256])$ for x, y dimensions and $min([2^{3 + depth}],\allowbreak[128])$ for z dimension.
The reasoning behind this was that resolution along the axial plane was often seen to be lower than in-plane resolution.
Patches were always randomly sampled from the whole scan on-the-fly.
In case of patch-size being larger than the whole scan, padding with reflecting border mode was performed along the required dimensions.

For the loss function, the generalized Dice loss was employed due to its robustness against label-inbalance \cite{Sudre2017}.
The volume-weighting factor $w_l$ was set to 1 for all classes, as otherwise an increase in false positives was observed for the less frequent classes.

As medical datasets are often small, heavy data augmentation was employed to increase generalizability of trained models \cite{Cicek2016}.
These augmentations comprise gaussian noise, rotation, scaling, as well as elastic deformation and were performed by a modified version of the batchgenerators framework \footnote{\url{https://github.com/MIC-DKFZ/batchgenerators}}.
While augmentation parameters are set globally for all tasks, the elastic deformation field is inversely scaled by the median voxel-spacing to ensure isotropic deformation in world coordinates.
Furthermore, augmentations were performed on-the-fly during training for maximal diversity of sampled augmentations.

\subsection{Inference}
A patch-wise approach is also used for inference.
Patch-sizes are clipped to $[128,512,512]$ due to GPU memory constraints.
Scans are first resampled to the median voxel-spacing of a given task, and patch-wise prediction is performed.
Afterwards, predictions are resampled to original voxel-spacing.
No further post-processing typically seen in biomedical imaging challenges such as CRFs is applied \cite{Dou2017}.
This is done to keep the model as simple as possible and avoid further hyperparameters.

\section{Results}
\begin{figure}[htbp]
  \centering
  \includegraphics[width = \textwidth]{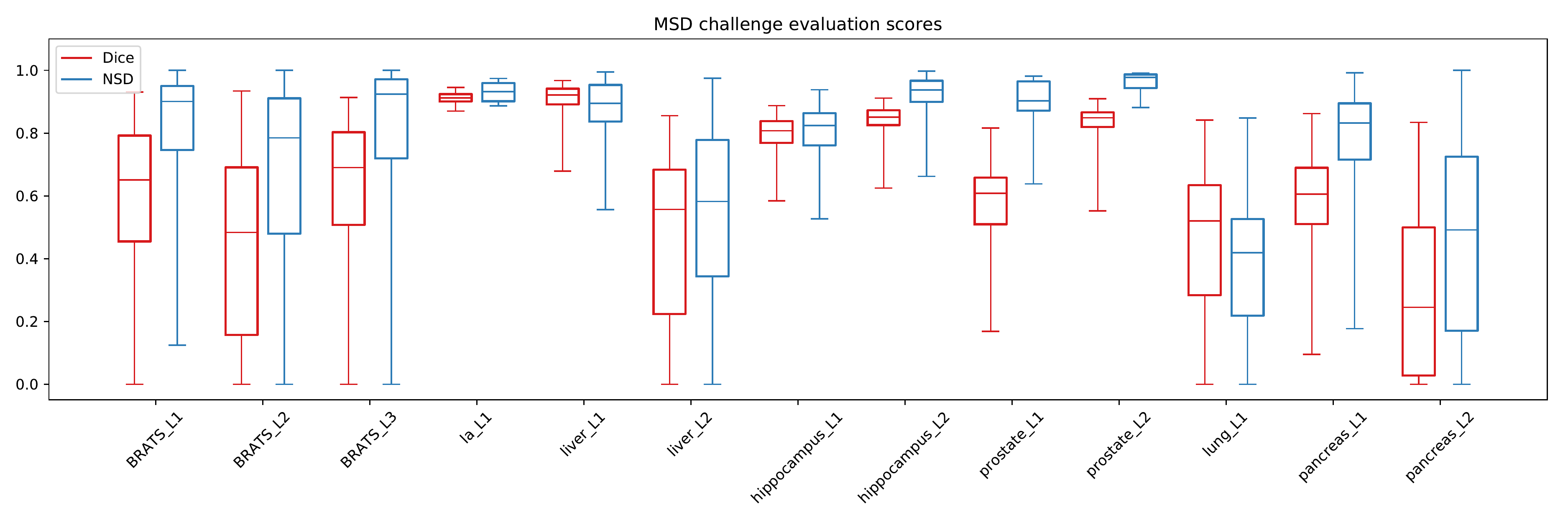}
  \caption{Evaluation metrics of the MSD challenge for the seven tasks released so far.
  Boxplot whiskers denote minimum as well as maximum value for the given metric.
  Boxes denote 25 and 75 percentiles, with the median depicted as a line inside the box.}
  \label{fig:evaluation metrics}
\end{figure}

For evaluation of the devised generalizable segmentation algorithm, the model was applied to the seven tasks released so far and the predictions submitted to the online evaluation portal of the MSD challenge.
The results are visualized in \cref{fig:evaluation metrics}.

\section{Discussion \& Conclusion}
Our generalizable segmentation algorithm designed for the MSD challenge was summarized in this paper.
When assessing preliminary results shown in \cref{fig:evaluation metrics}, it can be seen that the anisotropic model depth is suitable to handle scans with strongly varying voxel-spacings and scan shapes, as good results were achieved for the heart and prostate segmentation tasks.

Furthermore, the results achieved when trying to find small structures in large scans (liver, lung and pancreas tasks respectively) are not as good as in the other tasks.
Here, simultaneously low values for the Dice score and Normalized Surface Distance (NSD) indicate presence of false positives in off-site locations confirmed by visual inspection of predictions.
This is most likely caused by severe label inbalance, which make these segmentation tasks difficult to solve.
This is further exacerbated by random sampling of small patches and the used Dice loss, which is reliant on positive voxels being present in the ground truth of a patch \cite{Sudre2017}.

Nonetheless, the presented algorithm, due to its relative simplicity, provides a good basis for further development.
In future, we will work on integrating multi-scale approaches into the presented algorithm to increase performance on problems with severe label imbalance. Furthermore, integration of hyper-parameter searching techniques such as random search into our pipeline will be pursued to facilitate automatic, task-specific parameter-tuning.

%
%
%
%
\bibliography{./lit}

\begin{thebibliography}{10}
\providecommand{\url}[1]{\texttt{#1}}
\providecommand{\urlprefix}{URL }

\bibitem{Ayache2016}
Ayache, N., Duncan, J.: 20th anniversary of the medical image analysis journal
  (media). Medical Image Analysis  33,  1--3 (2016)

\bibitem{Cardoso2018}
Cardoso, M.J., Simpson, A., Ronneberger, O., Menze, B., van Ginneken, B.,
  Landman, B., Litjens, G., Farahani, K., Summers, R., Maier-Hein, L.,
  Kopp-Schneider, A., Bakas, S., Antonelli, M.: The medical segmentation
  decathlon challenge (7 2018), \url{http://medicaldecathlon.com/}

\bibitem{Chen2017}
Chen, H., Dou, Q., Yu, L., Qin, J., Heng, P.A.: Voxresnet: Deep voxelwise
  residual networks for brain segmentation from 3d mr images. NeuroImage
  (2017)

\bibitem{Cicek2016}
{\c{C}}i{\c{c}}ek, {\"O}., Abdulkadir, A., Lienkamp, S.S., Brox, T.,
  Ronneberger, O.: 3d u-net: learning dense volumetric segmentation from sparse
  annotation. In: International Conference on Medical Image Computing and
  Computer-Assisted Intervention. pp. 424--432. Springer (2016)

\bibitem{Dou2017}
Dou, Q., Yu, L., Chen, H., Jin, Y., Yang, X., Qin, J., Heng, P.A.: 3d deeply
  supervised network for automated segmentation of volumetric medical images.
  Medical image analysis  41,  40--54 (2017)

\bibitem{Isensee2018}
Isensee, F., Kickingereder, P., Wick, W., Bendszus, M., Maier{-}Hein, K.H.:
  Brain tumor segmentation and radiomics survival prediction: Contribution to
  the {BRATS} 2017 challenge. CoRR  abs/1802.10508 (2018),
  \url{http://arxiv.org/abs/1802.10508}

\bibitem{Kingma2014}
Kingma, D.P., Ba, J.: Adam: A method for stochastic optimization. arXiv
  preprint arXiv:1412.6980  (2014)

\bibitem{Maier-Hein*2018}
{Maier-Hein*}, L., {Eisenmann*}, M., {Reinke}, A., {Onogur}, S., {Stankovic},
  M., {Scholz}, P., {Arbel}, T., {Bogunovic}, H., {Bradley}, A.P., {Carass},
  A., {Feldmann}, C., {Frangi}, A.F., {Full}, P.M., {van Ginneken}, B.,
  {Hanbury}, A., {Honauer}, K., {Kozubek}, M., {Landman}, B.A., {M{\"a}rz}, K.,
  {Maier}, O., {Maier-Hein}, K., {Menze}, B.H., {M{\"u}ller}, H., {Neher},
  P.F., {Niessen}, W., {Rajpoot}, N., {Sharp}, G.C., {Sirinukunwattana}, K.,
  {Speidel}, S., {Stock}, C., {Stoyanov}, D., {Aziz Taha}, A., {van der
  Sommen}, F., {Wang}, C.W., {Weber}, M.A., {Zheng}, G., {Jannin*}, P.,
  {Kopp-Schneider*}, A.: Is the winner really the best? a critical analysis of
  common research practice in biomedical image analysis competitions. ArXiv
  e-prints  (Jun 2018)

\bibitem{Milletari2016}
Milletari, F., Navab, N., Ahmadi, S.A.: V-net: Fully convolutional neural
  networks for volumetric medical image segmentation. In: 3D Vision (3DV), 2016
  Fourth International Conference on. pp. 565--571. IEEE (2016)

\bibitem{Shen2017}
Shen, D., Wu, G., Suk, H.I.: Deep learning in medical image analysis. Annual
  Review of Biomedical Engineering  19(1),  221--248 (2017),
  \url{https://doi.org/10.1146/annurev-bioeng-071516-044442}, pMID: 28301734

\bibitem{Sudre2017}
Sudre, C.H., Li, W., Vercauteren, T., Ourselin, S., Cardoso, M.J.: Generalised
  dice overlap as a deep learning loss function for highly unbalanced
  segmentations. In: Deep Learning in Medical Image Analysis and Multimodal
  Learning for Clinical Decision Support, pp. 240--248. Springer (2017)

\bibitem{Ulyanov2016}
Ulyanov, D., Vedaldi, A., Lempitsky, V.S.: Instance normalization: The missing
  ingredient for fast stylization. CoRR  abs/1607.08022 (2016),
  \url{http://arxiv.org/abs/1607.08022}

\end{thebibliography}
\bibliographystyle{style/splncs03}

\end{document}